\documentclass[prb,superscriptaddress,twocolumn]{revtex4-2}
\usepackage{graphicx}
\usepackage{times,amsmath}
\usepackage{epsfig}
\usepackage{color}
\usepackage{graphicx}% Include figure files
\usepackage{dcolumn}% Align table columns on decimal point
\usepackage{bm}% bold math
\usepackage{bookmark}
\usepackage{tabularx}% bold math
\usepackage{hyperref}
\usepackage{multirow}
\usepackage{array,mathtools,amssymb,booktabs,makecell}
\hypersetup{colorlinks=true, citecolor=blue, filecolor=blue, linkcolor=blue, urlcolor=blue}

\usepackage{float}
\usepackage{adjustbox}
\usepackage{upgreek}
\usepackage{soul}
\makeatletter
\let\saved@includegraphics\includegraphics
\AtBeginDocument{\let\includegraphics\saved@includegraphics}

\makeatother

\begin{document}

\title{Strong electron correlations and ligand hybridization for altermagnetism}

\author{Byungkyun Kang}
\email[]{bkkang@utep.edu}
\affiliation{Department of Physics, 500 W University Ave, The University of Texas at El Paso, El Paso, Texas 79968, USA}

\author{Anderson Janotti}
\affiliation{Department of Materials Science and Engineering, University of Delaware, Newark, Delaware 19716, USA}
%\email{janotti@udel.edu}

\author{Dai Q. Ho}
\affiliation{Department of Materials Science and Engineering, University of Delaware, Newark, Delaware 19716, USA}
%\email{daiqho@udel.edu}

\author{Myoung-Hwan Kim}
\affiliation{Department of Physics and Astronomy, Texas Tech University, Lubbock, Texas 79409, USA}
%myounghwan.Kim@ttu.edu

\author{Chul Hong Park}
\affiliation{Quantum Matter Core-Facility and Research Center of Dielectric and Advanced Matter Physics, Pusan National University, Busan 46240, Republic of Korea}
%cpark@pusan.ac.kr

\author{Sangkook Choi}
\affiliation{School of Computational Sciences, Korea Institute for Advanced Study, Seoul 02455, Republic of Korea}
%sangkookchoi@kias.re.kr

\author{Mark R. Pederson}
\affiliation{Department of Physics, 500 W University Ave, The University of Texas at El Paso, El Paso, Texas 79968, USA}
%mrpederson@utep.edu

\author{Eunja Kim}
\email[]{ekim4@utep.edu}
\affiliation{Department of Physics, 500 W University Ave, The University of Texas at El Paso, El Paso, Texas 79968, USA}

%email
% Anderson Janotti   janotti@udel.edu
% Dai Q. Ho   daiqho@udel.edu
% Myoung-Hwan Kim   myounghwan.Kim@ttu.edu
% Chul Hong Park  cpark@pusan.ac.kr
% Sangkook Choi   sangkookchoi@kias.re.kr
% Mark R. Pederson   mrpederson@utep.edu
% Eunja Kim   ekim4@utep.edu

\begin{abstract}
Spin-band splitting is a hallmark of altermagnetism, intrinsically linked to magnetic ordering driven by electron correlations. However, recent inconsistencies in the detection of altermagnetism in strongly correlated altermagnet candidates have cast doubt on the robustness of this phenomenon and its dependence on many-body effects.
Here, \textcolor{black}{density functional theory combined with dynamical mean-field theory (DFT+DMFT)}, we dissect the electronic origins of altermagnetism in three prototypical candidates: MnF$_2$, MnTe, and RuO$_2$.
In MnF$_2$, we identify pronounced local electron correlations within Mn-3$d$ states and uncover a distinct Mott gap in the visible range. The strong correlations markedly localize the Mn-3$d$ electrons, leading to a narrowing of the spin-resolved bandwidth and, consequently, a suppression of spin-band splitting.
By contrast, MnTe provides an ideal platform for altermagnetism, exhibiting substantial local Mn-3$d$ magnetic moments due to the strong correlations and pronounced spin-band splitting, enabled by robust Mn-3$d$--Te-5$p$ orbital hybridization. 
RuO$_2$ manifests as a Pauli paramagnet with vanishing local moments, even in its antiferromagnetic phase. Nonetheless, it exhibits significant spin-band splitting, indicative of itinerant altermagnetic behavior.
Our results reveal that both strong local electron correlations and judicious ligand selection to promote orbital hybridization are key prerequisites to realizing altermagnetism in strongly correlated systems. These insights pave the way for the rational design and discovery of novel altermagnetic materials.

\end{abstract}

\maketitle

\subsection*{
Introduction.}
Altermagnetism emerges from the breaking of time-reversal symmetry in antiferromagnetic materials, resulting in the lifting Kramers degeneracy and the consequent splitting of spin bands, which can be experimentally detected via techniques such as angle-resolved photoemission spectroscopy (ARPES), Hall effect measurements, and neutron scattering~\cite{libor_natrev2022,olena_sciadv2024,libor_sciadv2020,rafael_prl2021,igor_pnas2021,makoto_ncom2019,zexin_nelec2022,ding_ncom2021,libor_prx2022,lee_prl2023,osumi_prb2024,krempasky_nature2024,xuhao_prl2025,sang_npjQM2025,lin_prb2020,sayantika_prx2024,yu_nature2024,sonka_ncomm2023,christopher_commat2026,tomas_nat2026,guowei_ncom2025}. 
However, recent studies have reported inconsistencies regarding the observation of altermagnetism. In RuO$_2$, both density functional theory (DFT) calculations and experiments, including ARPES, have indicated time-reversal symmetry breaking~\cite{ding_ncom2021,olena_sciadv2024}. Specifically, DFT results have found a significant local magnetic moment of 1.2 $\mu_{\mathrm{B}}$~\cite{basak_acta2024} and Hartree-Fock calculations yield 0.8 $\mu_{\mathrm{B}}$~\cite{Kyohoon_prb}. In contrast, polarized neutron diffraction experiments have reported a much smaller moment of 0.05 $\mu_{\mathrm{B}}$ and suggested itinerant antiferromagnetism in RuO$_2$~\cite{berlijn_prl2017}. Nonetheless, muon spectroscopy measurements and inelastic X-ray and neutron scattering have shown absence of magnetic moments~\cite{philipp_npjspin2024,george_cellrep2025}. Furthermore, longitudinal resistivity and Hall resistivity measurements reveal no evidence for a magnetic transition~\cite{xin_commat2025}. 
Spin-ARPES measurements show no evidence of spin-splitting~\cite{jiayu_prl2024}. 
These studies suggest that RuO$_2$ behaves as a non-magnetic material.
In MnF$_2$, theoretical studies based on DFT combined with dynamical mean-field theory (DMFT) predict a spin-band splitting of $\sim$ 0.7 eV~\cite{xuhao_prl2025}. 
In contrast, inelastic neutron scattering experiments with an energy resolution of $\sim$0.1 meV have not observed any signatures of altermagnetic magnon band splitting in MnF$_2$~\cite{morano_prl2025}.
Given that many candidate materials for altermagnetism are based on correlated transition-metal compounds, these discrepancies underscore the need for rigorous investigations into the influence of strong electron correlations, as well as other potentially hidden physical or chemical factors, on altermagnetic behavior.

\begin{figure*}[ht]
\centering
\includegraphics[width=0.9
\textwidth]{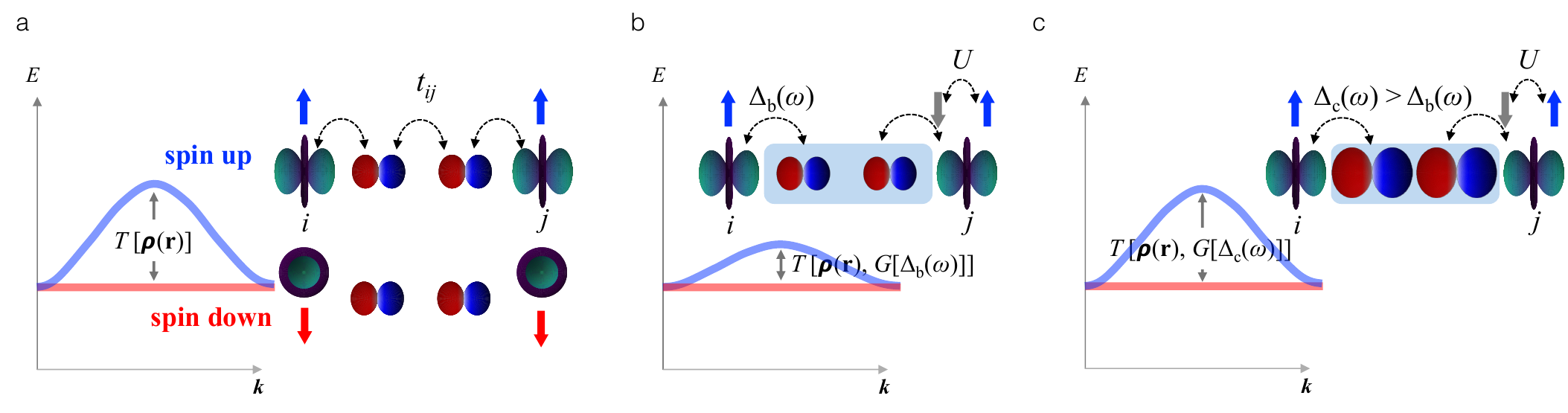}
\caption{\label{Fig_schematic}\
\textcolor{black}{\textbf{Schematic illustration of the impact of strong electron correlation and ligand hybridization on spin-band splitting.}} (\textbf{a}) Magnetic ordering via the superexchange pathway in an altermagnet. The spin-up band width is proportional to the electron kinetic energy, $T[\rho(\mathbf{r})]$, which depends on the hopping parameters $t_{ij}$ within DFT formalism. (\textbf{b}) Upon inclusion of many-body effects using DMFT, the hybridization $\Delta$ between correlated electrons and the conduction electron bath (indicated by the blue rectangle) is considered. The presence of Hubbard interaction $U$ leads to increased electron localization, resulting in a reduced spin-up band width associated with kinetic energy $T[\rho(\mathbf{r}), G[\Delta(\omega)]]$ within the DMFT framework. (\textbf{c}) An appropriate choice of a dispersive ligand with larger $\Delta$ enhances the kinetic energy, leading to an increased spin-up band width.
}
\end{figure*}

\textcolor{black}{A widely adopted framework for describing many-body electronic effects in solids is the single-band Hubbard model~\cite{antonio_rmp1996,gabriel_pt2004}. Its Hamiltonian is given by
\begin{equation} \label{eq:hubbard}
    H = - \sum_{\langle ij \rangle,\, \sigma} t_{ij} \left( {c}^{\dagger}_{i\sigma} {c}_{j\sigma} + \text{h.c.} \right)
    + U \sum_{i} {n}_{i\uparrow} {n}_{i\downarrow}
\end{equation}
Here, $\text{h.c.}$ denotes the Hermitian conjugate, and $t_{ij}$ represents the amplitude for an electron with spin $\sigma$ to hop between lattice sites $i$ and $j$. The term $U$ quantifies the local Coulomb repulsion experienced by two electrons occupying the same site $i$. The operators ${c}^{\dagger}_{i\sigma}$ and ${c}_{i\sigma}$ are the electron creation and annihilation operators, respectively, pertaining to site $i$ and spin $\sigma$. The density of electrons at site $i$ with spin $\sigma$ is ${n}_{i\sigma} = {c}^{\dagger}_{i\sigma} {c}_{i\sigma}$.
}

\textcolor{black}{DMFT solves the Hubbard model by mapping the interacting lattice problem onto a self-consistently defined Anderson impurity model. This approach treats local quantum fluctuations exactly, effectively capturing the evolution of the system’s many-body Green's function---which describes the propagation of an added or removed electron in energy space---according to
\begin{equation} \label{eq:Green}
G\left[\Delta(\omega)\right] = \sum_{\mathbf{k}} \left\{\omega - \Sigma[\Delta(\omega)] - t_{\mathbf{k}}\right\}^{-1}
\end{equation}
where $\Sigma$ is the self-energy, which captures interaction effects, and $\Delta(\omega)$ is the hybridization function encapsulating the dynamic coupling between the impurity and the electron bath. Here, $t_{\mathbf{k}}$ denotes the Fourier transform of the hopping matrix elements $t_{ij}$ for the original lattice~\cite{gabriel_pt2004,antonio_rmp1996}.
The combination of DMFT with DFT or GW methods has successfully captured strong electron correlation phenomena, such as Kondo effects~\cite{byung_usbte,kang2023dual,byung_ute2,kang_ndnio2,kang_ute2fs}, Hund's metal and Mott physics~\cite{kang_prb2026,kang_nqm2023,kang_nio, kang_fese}, and topological singularity-induced Mott-like self-energy~\cite{kang_tsme,kang_advsci2026}.
}
\textcolor{black}{As illustrated schematically in Fig.~\ref{Fig_schematic}a, in altermagnetic materials, the superexchange mechanism facilitates electron hopping between magnetic ions via intervening ligand orbitals~\cite{dai_arxiv2026}, where the bandwidth is proportional to the electronic kinetic energy $T[\rho(\textbf{r})]$ within the DFT formalism~\cite{gabriel_pt2004}.
When many-body effects are incorporated by solving the DMFT Green's function for the Hubbard model (Fig.\ref{Fig_schematic}b), electron correlations lead to a renormalization of the hopping parameter $t$. This accounts for increased electrons localization and the reduced bandwidth of the spin-up channel induced by the Hubbard interaction~\cite{wei_prb2017}. Such correlation effects are reflected in a reduced kinetic energy, expressed as $T[\rho(\textbf{r}),G[\Delta(\omega)]]$ within the DMFT framework~\cite{gabriel_pt2004}.
Importantly, the $G\left[\Delta(\omega)\right]$ depends explicitly on the hybridization function. By engineering ligand orbitals to be more dispersive, the hybridization strength can be enhanced, thereby increasing $\Delta(\omega)$. As a result, the effective hopping amplitude and the spin-up bandwidth are both increased, as shown schematically in Fig.~\ref{Fig_schematic}c.
}

\begin{figure*}[ht]
\centering
\includegraphics[width=1.0
\textwidth]{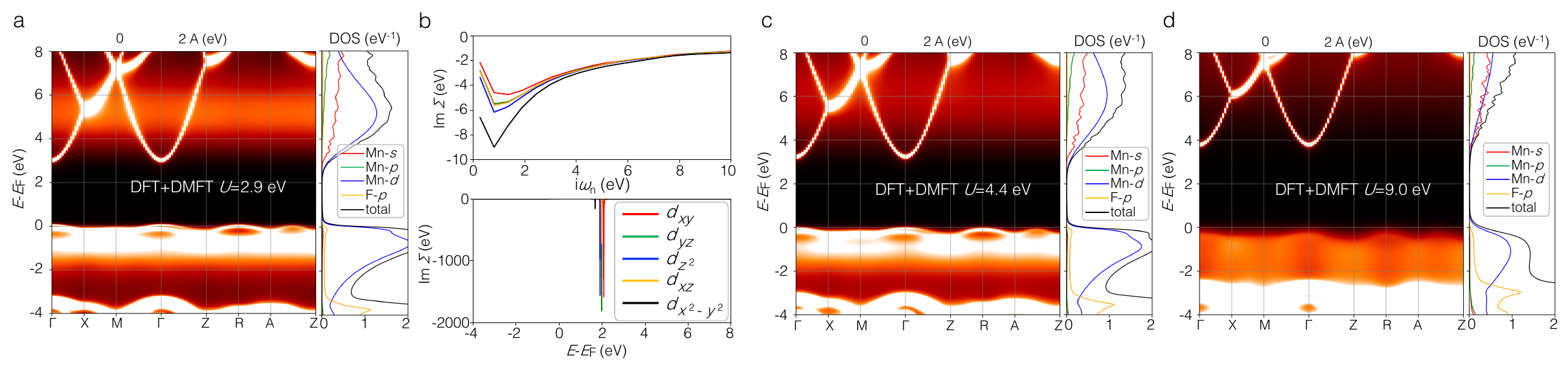}
\caption{\label{Fig_mnf2}\
\textcolor{black}{\textbf{Electronic structure of MnF$_2$.}}
Calculated spectral function and density of states (DOS) of MnF$_2$ using DFT+DMFT for $U$=2.9 eV (\textbf{a}), $U$ = 4.4 eV (\textbf{c}), and $U$ = 9.0 eV (\textbf{d}).
(\textbf{b}) Self-energy of Mn-3$d$ orbitals in MnF$_2$ calculated using DFT+DMFT with $U = 2.9$ eV.
In \textbf{a},\textbf{c} and \textbf{d}, the Mn-3$d$ (4$s$) DOS are scaled by factors of 4 for visibility.
All calculations were performed at an electronic temperature of 1000 K in the paramagnetic phase.
}
\end{figure*}

\textcolor{black}{
In this study, we investigate the electronic structures of the altermagnet candidates MnF$_2$, MnTe, and RuO$_2$ using the DFT+DMFT framework~\cite{choi2019comdmft}. Our calculations reveal that MnF$_2$ is a strongly correlated material that exhibits a band gap within the visible energy range. Furthermore, we extend our analysis to both MnF$_2$ and MnTe in order to elucidate the interplay between strong electron correlations and ligand hybridization in determining the spin-band splitting.
Our results show that strong electron correlations in MnF$_2$ tend to suppress spin-band splitting due to enhanced electron localization. In contrast, MnTe, which exhibits both strong correlation effects and pronounced ligand hybridization, displays substantial spin-band splitting. For comparison, RuO$_2$ shows no clear indications of strong electronic correlations. These results underscore the crucial role of ligand hybridization in enabling altermagnetic behavior in strongly correlated systems.
}

\begin{figure*}[ht]
\centering
\includegraphics[width=0.9
\textwidth]{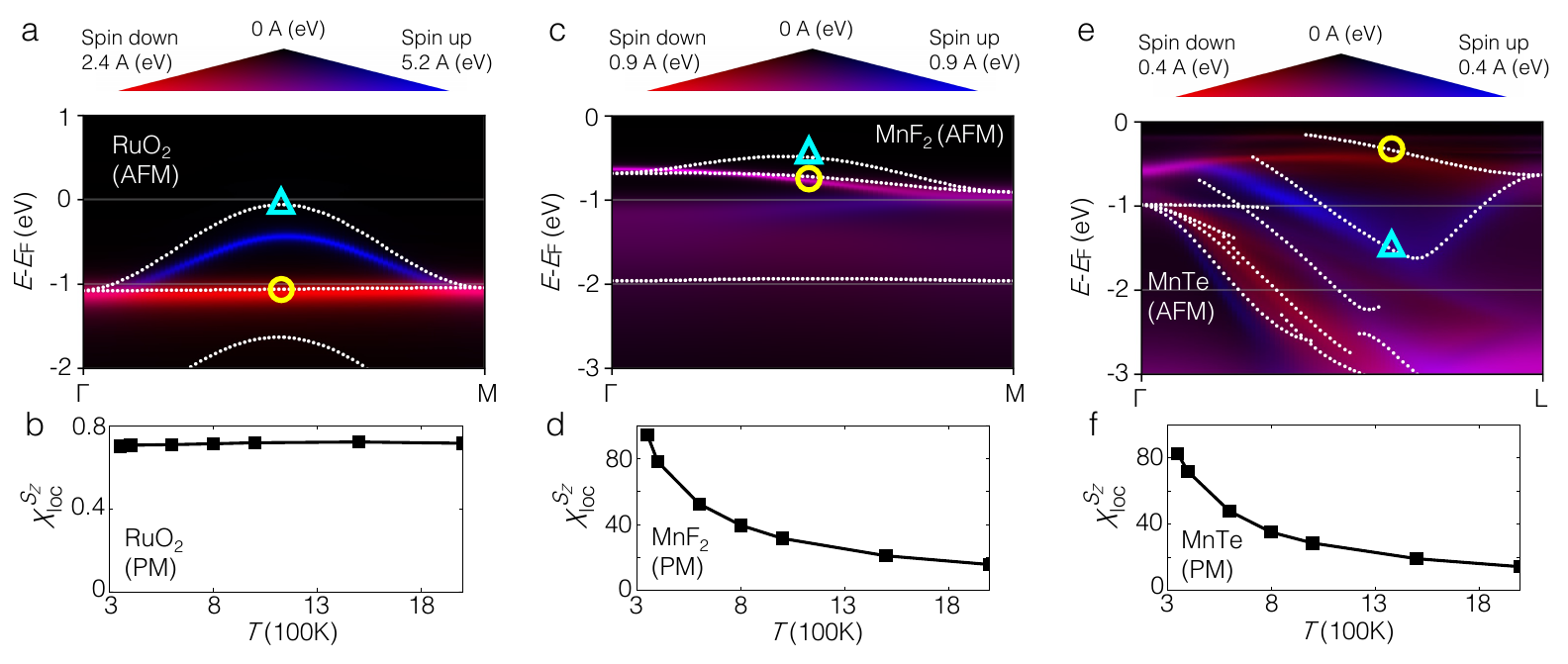}
\caption{\label{Fig_spinband}\
\textbf{Spin-band splittings.}
Calculated spin-resolved spectral functions, obtained via DFT+DMFT in the antiferromagnetic (AFM) phase are shown for (\textbf{a}) RuO$_2$ (\textcolor{black}{$T$=300 K}), (\textbf{c}) MnF$_2$ (\textcolor{black}{$T$=40 K}), and (\textbf{e}) MnTe (\textcolor{black}{$T$=300 K}). 
The corresponding single-particle bands from DFT+$U$ (dotted white lines; blue triangles for spin-up, yellow circles for spin-down) are overlaid to facilitate the identification of many-body effects present in the DFT+DMFT results.
Calculated local spin susceptibility,
$\chi_{\mathrm{loc}}^{S_z}$ from DFT+DMFT in the paramagnetic (PM) phase for (\textbf{b}) RuO$_2$, (\textbf{d}) MnF$_2$, and (\textbf{f}) MnTe.
}
\end{figure*}

\subsection*{Results and Discussion}
\subsubsection*{Mott insulator MnF$_2$.}

\textcolor{black}{We conducted charge self-consistent DFT+DMFT calculations using $U$ of 2.9 eV, 4.4 eV and 9.0 eV, as shown in Fig.~\ref{Fig_mnf2}a,c,d. The resulting spectral functions exhibit fundamental band gaps of $\sim$2 eV ($U = 2.9$ eV), $\sim$3 eV ($U = 4.4$ eV), and $\sim$4 eV ($U = 9.0$ eV). The valence bands are predominantly composed of Mn-3$d$ lower Hubbard bands (LHB), which exhibit increased coherence with decreasing $U$. The conduction bands consist of incoherent Mn-3$d$ upper Hubbard bands (UHB) as well as highly dispersive Mn-4$s$ states. As $U$ is reduced, both Mn-3$d$ and Mn-4$s$ states shift toward the valence band edge. This behavior indicates that the energy levels of the Mn-3$d$ LHB and UHB are mainly governed by the $U$ among Mn-3$d$ electrons. In contrast, the position of the Mn-4$s$ bands is modulated by the Mn-4$d$ UHB.}

\textcolor{black}{
Although the band gap of approximately 2 eV obtained from DFT+DMFT calculation with $U = 2.9$ eV is in good agreement with experimental observations of optical absorption features in the visible range near 2.4 eV~\cite{li_jpcs2009,kwon_jkms2007,caird_jap1971,ignacio_jpcm2007,taiju_prb1991,hernandez_prl2007}—values typically associated with weakly correlated band semiconductors—the electronic structure of MnF$_2$ exhibits clear signatures of a Mott insulator, arising from strong local electron correlations within the Mn-3$d$ orbitals.
As illustrated in Fig.~\ref{Fig_mnf2}b, the divergent local impurity self-energy of Mn-$d$ in MnF$_2$ is analogous to that observed in Mott insulators such as NiO~\cite{kangfgwedmft} and FeSe~\cite{kang_fese}, suggesting that MnF$_2$ is governed by Mott physics. This is further supported by the observation of both the LHB and UHB in all five half-filled Mn-3$d$ orbitals of MnF$_2$, since the calculated Mn-3$d$ occupancies are close to 1.0.
}

\begin{figure*}[ht]
\centering
\includegraphics[width=0.9
\textwidth]{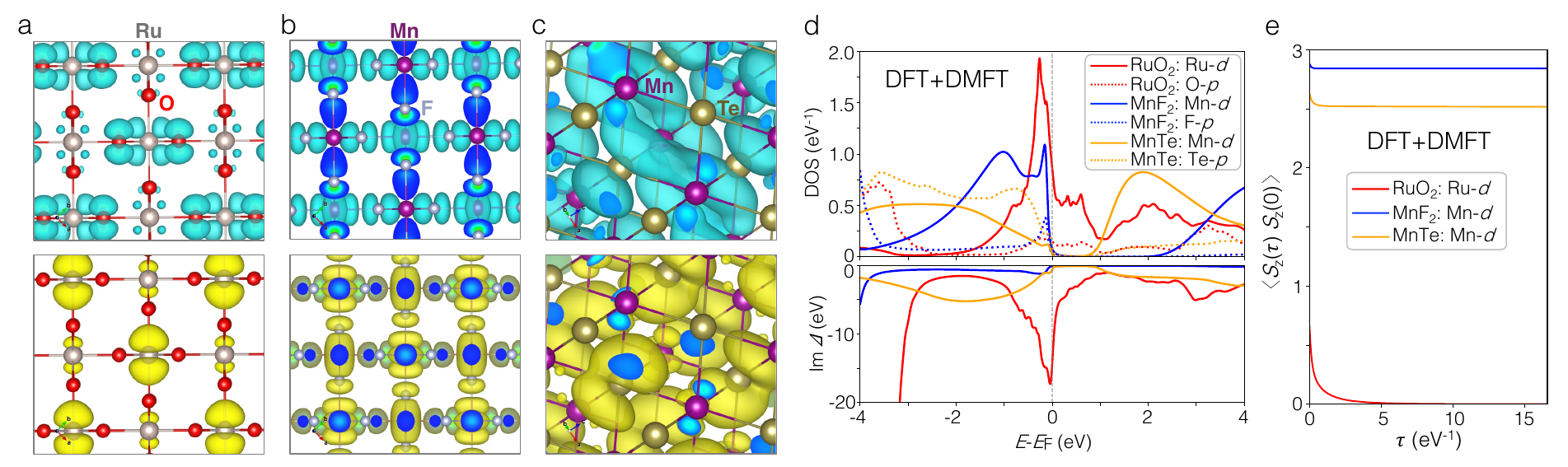}
\caption{\label{Fig_hyb}\
\textbf{Ligand hybridization.}
The squared wavefunctions of the spin-up (blue, upper panel) and spin-down (yellow, lower panel) states correspond to the band and \textbf{k}-point selections indicated by blue triangles (spin-up) and yellow circles (spin-down) in Fig.~\ref{Fig_spinband} for (\textbf{a}) RuO$_2$, (\textbf{b}) MnF$_2$, and (\textbf{c}) MnTe.
All isosurfaces are obtained from DFT+$U$ calculations and plotted with a saturation level of 1$\times$10$^{-9}$.
(\textbf{d}) Orbital-projected density of states (upper panel) and hybridization function ($\Delta$) (lower panel) calculated in the paramagnetic phase at 350 K using DFT+DMFT.
(\textbf{e}) Local spin moment correlation functions in the paramagnetic phase at 350 K, as obtained from DFT+DMFT.
}
\end{figure*}

\subsubsection*{
Impact of electron correlations and ligand hybridization on altermagnetism.}
We calculated the spin-band splitting in the altermagnetic candidates RuO$_2$, MnF$_2$, and MnTe using both DFT+DMFT, as implemented in ComDMFT~\cite{choi2019comdmft}, and DFT+$U$, as implemented in VASP~\cite{kresse1994ab,kresse1996efficient,blochl1994projector,kresse1999ultrasoft}. The $U$ was set to 1.7 eV~\cite{Kyohoon_prb} for RuO$_2$, and to 2.9 eV for both MnF$_2$ and MnTe. The $U$ = 2.9 eV is calculated from cRPA~\cite{aryasetiawan2004frequency,aryasetiawan_calculations_2006} for MnTe, as implemented in LQSGW+DMFT~\cite{choi2019comdmft}. For MnF$_2$, DFT+DMFT calculation with $U$ = 2.9 eV yield a spectral function \textcolor{black}{exhibiting a visible range band gap, in good agreement with experimental observations~\cite{li_jpcs2009,kwon_jkms2007,caird_jap1971,ignacio_jpcm2007,taiju_prb1991,hernandez_prl2007}.} For MnTe, we also conducted DFT+DMFT calculations using $U$ = 5 eV~\cite{lee_prl2023}, which produces a qualitatively similar electronic structure with an increased band gap.

As MnF$_2$ was identified as a strongly correlated material in the previous section, its local spin susceptibility--defined as
$\chi_{\mathrm{loc}}^{S_z}=\int_0^\beta d\tau\langle S_z(\tau)S_z(0)\rangle$ 
and obtained by DFT+DMFT--exhibits Curie-Weiss behavior in the paramagnetic phase, as shown in Fig.~\ref{Fig_spinband}d.
This indicates the presence of a significant local magnetic moment of the Mn-3$d$ electrons, which may play an important role in altermagnetic behavior. However, our analysis reveals that strong electron correlations suppress the spin-band splitting. 
As depicted in Fig.~\ref{Fig_spinband}c, the DFT+$U$ calculations (white dashed lines) for MnF$_2$ in the antiferromagnetic (AFM) phase display a spin-band splitting of $\sim$0.3 eV along the $\Gamma$--M high-symmetry direction, resulting from the breaking of time-reversal symmetry.
\textcolor{black}{Within the DFT+DMFT framework, which explicitly incorporates many-body effects (as described by Eqs.~\eqref{eq:hubbard}, \eqref{eq:Green} and illustrated in Fig.~\ref{Fig_schematic}a,b),} the spin-band splitting observed in the spin-resolved spectral function at 40 K in the AFM phase is found to be negligible. 
\textcolor{black}{This suppression arises from the strong electron correlations that localize Mn-3$d$ electrons (decrease in the kinetic energy) leading to reduction of the spin-up bandwidth, as reflected in the Green's function and schematically depicted in Fig.~\ref{Fig_schematic}b. The strong electron localization is further corroborated by the substantial self-energy (Fig.~\ref{Fig_mnf2}b), Curie-Weiss magnetic susceptibility (Fig.~\ref{Fig_spinband}d), significant local magnetic moment (Fig.~\ref{Fig_hyb}e), and the emergence of incoherent Hubbard bands (Fig.~\ref{Fig_mnf2}a) in the paramagnetic phase.
}

\begin{figure*}[ht]
\centering
\includegraphics[width=1.0 \textwidth]{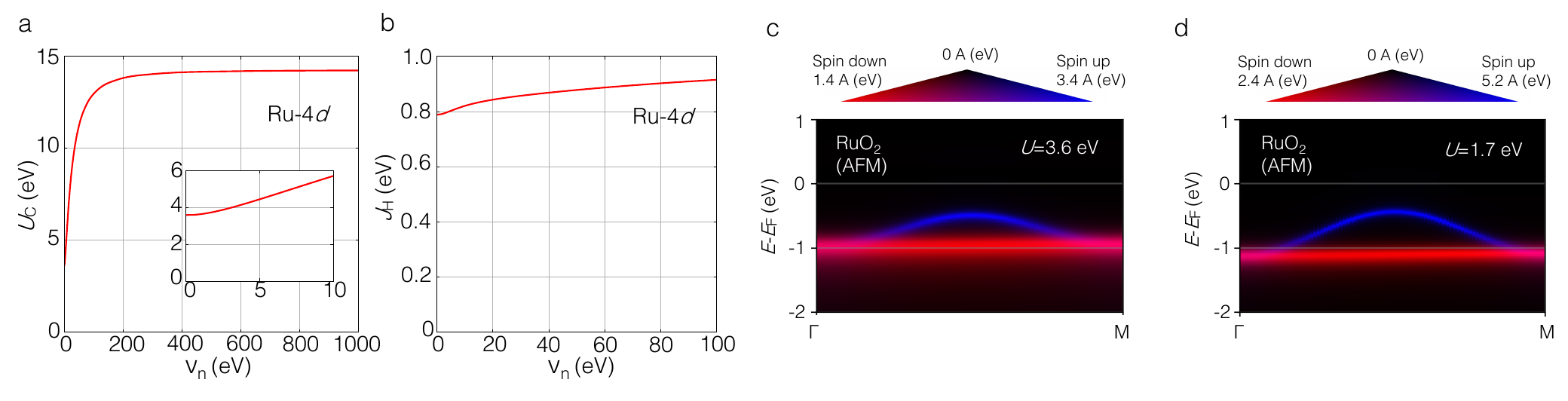}
\caption{\label{Fig_ruo2} 
\textbf{A larger $U$ of RuO$_2$.}
\textbf{a}, Coulomb interaction $U_{\textrm{C}}$ 
\textbf{b}, Exchange interaction $J_{\textrm{H}}$. 
Spectral functions from DFT+DMFT in the antiferromagnetic (AFM) phase for RuO$_2$ using $U$ = 3.6 eV (\textbf{c}) and $U$=1.7 eV (\textbf{d}),
}
\end{figure*}

\textcolor{black}{Our DFT+$U$ result is consistent with the theoretically predicted sizable spin-band splitting~\cite{xuhao_prl2025,sayantika_prx2024}. However, inelastic neutron scattering (INS) experiments have yielded conflicting results regarding the chirality of the magnon bands in MnF$_2$. 
Faure et al.~\cite{quentin_2025} reported a maximum magnon band splitting of approximately 0.2 meV, attributing this observation to altermagnetism. In contrast, Morano et al.~\cite{morano_prl2025}--with an energy resolution of $\sim$0.1 meV--did not observe any signatures of altermagnetic magnon band splitting, and thus argued for the absence of altermagnetism.
Although the nearly vanishing spin-band splitting obtained from the spin-resolved spectral function within DFT+DMFT is not directly comparable to the spin excitations measured by INS, due to the disparity in energy scales, it is important to note that strong electron correlations can significantly suppress the spin-band width. This suppression may, in turn, make experimental detection of magnon band splitting challenging.
}

For MnTe, the Curie-Weiss behavior of $\chi_{\mathrm{loc}}^{S_z}$, as displayed in Fig.~\ref{Fig_spinband}f, indicates strong electron correlations and the presence of local magnetic moments, similar to what is observed in MnF$_2$. However, in contrast to MnF$_2$, both DFT+DMFT and DFT+$U$ calculations show a significant spin-band splitting of $\sim$1 eV in MnTe (Fig.~\ref{Fig_spinband}e), consistent with previous reports~\cite{mazin_prb2023,ruben_npjspin2024,lee_prl2023,osumi_prb2024,krempasky_nature2024,zheyuan_prl2024}.

To investigate the distinct behaviors observed in RuO$_2$, MnF$_2$, and MnTe, we computed the local spin moment correlation functions within DFT+DMFT, defined as $\chi_{S_Z}(\tau)=\langle S_z(\tau)S_z(0)\rangle$, which are presented in Fig.\ref{Fig_hyb}e. These correlation functions serve as a diagnostic for evaluating the degree of magnetic moment localization~\cite{belozerow_prb2023}.
The $\chi_{S_Z}(\tau)$ at $\tau\xrightarrow{} \beta/2$ exhibits a vanishing local magnetic moment for RuO$_2$, whereas both MnF$_2$ and MnTe display substantial and comparable frozen local magnetic moments. This suggests that the degree of electron correlation is similarly strong in both MnF$_2$ and MnTe. 
However, as shown in Fig.\ref{Fig_hyb}d, the hybridization function ($\Delta$, \textcolor{black}{see Eqs.~\eqref{eq:Green}}) of Mn-3$d$ in MnTe is significantly greater than in MnF$_2$ within the energy window of -3 eV $<$ $E-E_{\textrm{F}}$ $<$ 0, where prominent spin-band splitting is observed. In this region, the Te-5$p$ DOS in MnTe is considerable, in contrast to the negligible F-2$p$ DOS in MnF$_2$, indicating substantial hybridization between Mn-3$d$ and Te-5$p$ orbitals.
The enhanced hybridization in MnTe is further evidenced by the wavefunctions of the spin-split bands obtained from DFT+$U$ calculations, as depicted in Fig.~\ref{Fig_hyb}a--c. 
In RuO$_2$, pronounced antibonding between O-2$p$ and Ru-4$d$ orbitals is observed in the dispersive spin-up band along the $\Gamma$–M path, while the spin-down band shows no such interaction. This behavior arises from the selective hybridization of magnetic Ru-4$d$ orbitals with non-magnetic ligand O-2$p$ orbitals in two tilted sublattices of the rutile unit cell, which breaks the half-unit cell translation and time-reversal symmetries~\cite{olena_sciadv2024,libor_sciadv2020}.
Similarly, the small spin-band splitting in MnF$_2$ arises from antibonding hybridization between Mn-3$d$ and F-2$p$ states on both spin-up and spin-down bands. 
However, the extent of hybridization between the Mn-3$d$ and Te-5$p$ orbitals in MnTe is significantly greater, owing to the bonding character present in the spin-up band. This enhanced hybridization leads to a pronounced dispersion of the spin-up band and, consequently, a substantial spin-band splitting in MnTe.
These findings indicate that although both MnTe and MnF$_2$ are strongly correlated systems with local magnetic moments, which are essential for altermagnetism, it is the substantial hybridization between Mn-3$d$ and Te-5$p$ orbitals unique to MnTe that primarily drives the spin-band splitting.

\subsubsection*{
\textcolor{black}{Itinerant altermagnetic behavior of RuO$_2$.}}

\textcolor{black}{For RuO$_2$, a material that has generated considerable debate regarding its potential as an altermagnetic candidate, we investigate the nature of spin-band splitting in the AFM phase. As illustrated in Fig.\ref{Fig_spinband}a, our DFT+$U$ calculations reveal a substantial spin-band splitting of approximately 0.9 eV. 
When many-body effects are explicitly taken into account using DFT+DMFT with $U$ = 1.7 eV, the spin-band splitting is slightly reduced to 0.7 eV, highlighting the impact of electron localization. To further investigate the influence of the $U$ on electron localization, $U$ was increased to 3.6 eV, as estimated by the cRPA method (see Fig.~\ref{Fig_ruo2}a,b). Notably, this larger $U$ value further decreases the spin-band splitting to 0.5 eV, indicative of enhanced electron localization. However, the overall suppression of the spin-band splitting remains minor. 
The persistence of significant spin-band splitting at both values of $U$ is consistent with previous theoretical studies reporting sizable spin-band splitting~\cite{ding_ncom2021,olena_sciadv2024}, suggesting that electron correlation effects play only a limited role in determining the spin-band splitting in RuO$_2$.
}

\textcolor{black}{Despite the presence of spin-band splitting, our DFT+DMFT calculations reveal that the local magnetic moment in the AFM phase remains nearly zero for both small and large values of $U$.
Nevertheless, the calculations stably converge with the sizable spin-band splitting, which is indicative of itinerant antiferromagnetism~\cite{berlijn_prl2017,zhu_prl2019}. Additionally, the calculated local spin susceptibility in the paramagnetic phase ($\chi_{\mathrm{loc}}^{S_z}$) remains nearly constant, a hallmark of Pauli paramagnetic behavior (see Fig.\ref{Fig_spinband}b). These results collectively suggest that the AFM phase in RuO$_2$ is likely unstable, consistent with recent experimental reports that observe no signatures of magnetism in RuO$_2$~\cite{philipp_npjspin2024,xin_commat2025,george_cellrep2025,jiabin_prb2025,zheyu_prx2025}.
}

\textcolor{black}{While the nature of magnetism in RuO$_2$ remains a subject of ongoing controversy, our findings suggest that RuO$_2$ is unlikely to exhibit conventional altermagnetic order, primarily due to the absence of a well-defined local magnetic moment—typically a signature of strong electron correlation. However, it should be noted that, as an itinerant antiferromagnet, RuO$_2$ retains the potential for spin-band splitting, thereby leaving open the possibility for an unconventional, itinerant altermagnetic state.
}

\subsection*{Conclusion.} 

In summary, our analysis establishes MnF$_2$ as a prototypical strongly correlated insulator with a visible-range Mott gap. Notably, the pronounced electronic correlations in MnF$_2$ act to suppress dispersive spin-band splitting. In contrast, MnTe exhibits the coexistence of strong electron correlations and substantial ligand hybridization, giving rise to significant spin-band splitting, while RuO$_2$ remains largely uncorrelated. Collectively, these results highlight the essential role of ligand hybridization in stabilizing altermagnetic behavior in strongly correlated systems.

\subsection*{\textcolor{black}{Methods}}

\noindent \textbf{DFT+DMFT calculation\\}

We used experimental lattice parameters for all compounds studied. For tetragonal MnF$_2$ (space group $P4_2/mnm$, No.\ 136), the lattice constants are $a = b = 4.87$ {\AA} and $c = 3.31$ {\AA}~\cite{griffel1950preparation}. For tetragonal RuO$_2$ (space group $P4_2/mnm$, No.\ 136), we used $a = b = 4.52$ {\AA} and $c = 3.12$ {\AA}~\cite{yin_ncomm2022}. For hexagonal MnTe (space group $P6_3/mmc$, No.\ 194), the lattice parameters are $a = b = 4.15$ {\AA} and $c = 6.71$ {\AA}~\cite{szuszkiewicz_prb2006}.

We performed charge-self-consistent DFT+DMFT calculations utilizing the COMSUITE package~\cite{choi2019comdmft}. The local Coulomb interactions for the correlated $d$-shell were parameterized using the Slater integrals: $F^0 = 2.9$, $4.4$, and $9.0$~eV, $F^2 = 7.7$~eV, and $F^4 = 6.2$~eV for MnF$_2$; and $F^0 = 2.9$, and $5.0$~eV, $F^2 = 7.7$~eV, and $F^4 = 6.2$~eV for MnTe, which define the interaction tensor for the impurity problem. The values $F^0 = 2.9$ eV, $F^2 = 7.7$ eV, and $F^4 = 6.2$ eV used for MnTe were obtained from constrained Random Phase Approximation (cRPA) calculations~\cite{aryasetiawan2004frequency,aryasetiawan_calculations_2006} as implemented in LQSGW+DMFT~\cite{choi2019comdmft}. Compared to the 
DFT+DMFT database~\cite{database}, which reports $F^0 = 9.0$~eV, $F^2 = 9.8$~eV, and $F^4 = 6.1$~eV, the present values for the Mn-$3d$ shell are notably lower. For RuO$_2$, the Slater integrals for Ru-4$d$ were set to $F^0 = 1.7$~eV, $F^2 = 3.5$ eV, and $F^4 = 2.9$ eV, consistent with previous optimization~\cite{Kyohoon_prb}. For the double-counting correction, we adopted the nominal double-counting scheme~\cite{haule2010dynamical,kristjan_prb2014} with a $d$-shell occupancy fixed at 5.0 for MnF$_2$ and MnTe and 4.0 for RuO$_2$.

\bigskip

\noindent \textbf{LQSGW+DMFT calculation\\}

We employed linearized quasiparticle self-consistent GW combined with DMFT (LQSGW+DMFT)\cite{choi2019comdmft}. This scheme is a simplified variant of fully self-consistent GW + extended dynamical mean-field theory (FGW+EDMFT) and does not achieve full self-consistency in the fermionic/bosonic sectors\cite{sun2002extended,biermann_prl2003,nilsson2017multitier,kangfgwedmft,kang_fese}. We construct the LQSGW quasiparticle Hamiltonian and  determine the Slater integrals, based on a constrained-RPA calculation~\cite{aryasetiawan2004frequency,aryasetiawan_calculations_2006}.

\bigskip
\noindent \textbf{DFT+$U$ calculation\\}

Electronic structure calculations and geometry optimizations were carried out utilizing the projector augmented-wave (PAW) method, as implemented in the Vienna \textit{Ab-initio} Simulation Package (\textsc{VASP})~\cite{blochl1994projector, kresse1996efficient, kresse1999ultrasoft}. Exchange-correlation interactions were described using the Generalized Gradient Approximation (GGA) with the Perdew-Burke-Ernzerhof (PBE) functional~\cite{perdew1996generalized}. To capture the strong electron correlation effects in the Mn-3$d$ and Ru-4$d$ orbitals, a Hubbard $U$ correction was included. The $U$ value was set to 1.7 eV for RuO$_2$ and 2.9 eV for both MnF$_2$ and MnTe, in accordance with the Slater integral $F^0$ parameters employed in our DFT+DMFT calculations. A plane-wave energy cutoff of 520 eV was uniformly applied in all calculations. Structural relaxations were performed until the residual forces on each atom were less than 0.03 eV/\AA. Brillouin zone integration was performed using $\Gamma$-centered $k$-point meshes of $6 \times 6 \times 4$ for MnTe and $5 \times 5 \times 8$ for both MnF$_2$ and RuO$_2$.
%~\cite{PhysRevLett.129.185701}

\section*{Acknowledgments}  
This research was performed using funding received from the U.S. Department of Energy, Office of Nuclear Energy’s Nuclear Energy University Program (NEUP).
We acknowledge the High Performance Computing Center (HPCC) at Texas Tech University for providing computational resources that have contributed to the research results reported within this paper.
C. H. P. acknowledges the support by the National Research Foundation of Korea (NRF) grant (Grant No. NRF-2022R1A2C1005548) and the Ministry of Education (grant No. 2021R1A6C101A429). S. C. was supported by a KIAS Individual
Grant (CG090601) at Korea Institute for Advanced Study

\bigskip
\textbf{Competing Interests} The authors declare no competing interests.

\bigskip
\textbf{Data availability} The data that support the ﬁndings of this study are available from the corresponding
authors upon reasonable request.

\bigskip
\textbf{Author contributions} 
B.K. and E.K. designed the project. B.K. performed the DFT+DMFT, LQSGW+DMFT and DFT+$U$  calculations and conducted the data analysis. 
All authors wrote the manuscript, discussed the results, and commented on the paper.
\bibliography{ref}

\end{document}